\documentclass[aps,amssymb,prl,showpacs,twocolumn ]{revtex4} \fussy

\usepackage{graphicx}
\usepackage{dcolumn}

\newcommand{\figwidth}{0.45\textwidth}

\begin{document}

\title{
Complex  response and
 polariton-like dispersion splitting in periodic metal nanoparticle chains}

\author{A. Femius Koenderink}\email{koenderink@amolf.nl}
\author{Albert Polman}
\affiliation{FOM Institute for Atomic and
Molecular Physics AMOLF, Center for Nanophotonics, Kruislaan 407,
1098 SJ Amsterdam, The Netherlands}

\date{Submitted to Phys. Rev., December 6, 2005.}

\begin{abstract}
We show that guiding of optical signals in  chains of metal
nanoparticles is subject to a
 surprisingly complex dispersion relation. Retardation causes the dispersion relation to split in
 two anticrossing branches, as common for polaritons. While huge radiation losses occur above the
 light line, just below the light line the micron-sized loss lengths are much longer than expected.
 The anticrossing allows to create highly localized energy distributions in  finite arrays that
 can be tuned via the illumination wavelength. Our results apply to all linear chains of coupled resonant scatterers.
\end{abstract}


 \maketitle

Ordered arrays of optically driven metal nanospheres may be used
to transport optical signals in structures that are much smaller
than the wavelength of light
~\cite{atwaterWGproposal,atwaternatmat,quinten,maierlossapl,maierextinction}.
Coherent   transport occurs for frequencies close to the plasmon
resonance of a single  nanosphere, and is mediated by
dipole-dipole coupling between resonant particles on a chain. The
transverse width of the guiding structure is only limited by the
particle diameter, which is an order of magnitude smaller than the
optical wavelength. Metal nanoparticle arrays may therefore allow
to overcome current limitations in the miniaturization of
integrated  optical devices. Although several experimental and
theoretical
studies~\cite{quinten,atwaterWGproposal,maierextinction,maierlossapl,atwaternatmat,park,citrin,weber}
confirm that coupling effects occur in metal nanosphere arrays,
no complete picture for the dispersion relation and propagation
loss has emerged.

Recent studies indicate that radiation damping, retardation
effects and long-range coupling, dramatically affect the loss and
dispersion~\cite{weber,citrin}. This necessitates a complete
revision of the original theoretical
studies~\cite{atwaterWGproposal,maierextinction,maierlossapl,park}.
In this paper we calculate the dispersion and loss including all
these effects for finite and infinite arrays. Remarkably,
retardation causes the dispersion to  split into two anticrossing
branches for transverse modes. Modes with wave vectors slightly
larger than those   in the embedding medium have the lowest loss,
and allow transport over micrometer distances, well in excess of
previous estimates~\cite{maierlossapl}. We further show that the
anticrossing gives rise to a strongly localized optical response
in finite arrays that is very sensitive to the incident wavelength
around the anticrossing range.  As our calculations are not
specific to metallic nanoparticles, our results apply in general
to periodic arrays of coupled resonant dipoles, in nano-optics as
well as in, e.g., atomic arrays interacting with
radiation~\cite{pedrodevries,atomlattice,robicheaux2}.

 We consider finite and infinite arrays of equally spaced metal
 nanospheres of radius $a$,   spaced by a center-to-center distance $d$.
 A coupled  point-dipole approximation is well suited to describe
 the electromagnetic response of such chains~\cite{park,weber,citrin,robicheaux}.
 We use the electric field generated by a single  dipole $\mathbf{p}  e^{-i\omega t}$, oscillating  with frequency $\omega$
\begin{eqnarray}\mathbf{E}(\mathbf p,
 \mathbf{r},\omega) & =&  \frac{1}{4\pi \epsilon }\left[\left(1-\frac{i\omega
 r}{v}\right)\frac{3(\hat{\mathbf{r}}\cdot\mathbf{p})\hat{\mathbf{r}}-\mathbf{p}}{r^3}\right.
 \nonumber\\ &&\left.
 +
 \frac{\omega^2}{v^2}\frac{\mathbf{p}-(\hat{\mathbf{r}}\cdot\mathbf{p})\hat{\mathbf{r}}}{r}\right]e^{i\omega
 r/v}\label{eq:dipolefield}
\end{eqnarray} where $\hat{\mathbf{r}}$ is the unit vector pointing from the
dipole to the field point at distance $r$,
$\epsilon=n^2\epsilon_0$ is the permittivity of the homogeneous
medium in which the dipole is embedded, and $v=c/n$ is the
corresponding speed of light. This form of the electric field
fully takes into account retardation effects. The induced dipole
moment on a nanosphere equals its polarizability $\alpha(\omega)$
times the local electric field, which is composed of the driving
 field $\mathbf{E}^{\rm{(ext)}}$ and the fields of all the other nanospheres. For the $n^{\rm th}$ nanosphere in a
linear chain pointing along $\hat{\mathbf{r}}$ we find
\begin{eqnarray}
\mathbf{p}_n&&=\alpha(\omega)\left\{\mathbf{E}^{(\rm{ext})}_n
+\frac{1}{4\pi\epsilon} \sum_{m\neq
n}\left[\left(1-\frac{i\omega|n-m|d}{v}\right)\right.\right.\nonumber\\
&& \left.\left. \cdot
\frac{3(\hat{\mathbf{r}}\cdot\mathbf{p_m})\hat{\mathbf{r}}-\mathbf{p_m}}{|n-m|^3d^3}
+\frac{\omega^2}{v^2}\frac{\mathbf{p_m}-(\hat{\mathbf{r}}\cdot\mathbf{p_m})\hat{\mathbf{r}}}{|n-m|d}\right]
e^{i\omega|n-m|d/v}\right\}. \label{eq:coupled}
\end{eqnarray}
 For   $N$ dipoles, Eq.~(\ref{eq:coupled}) represents  $N$ inhomogeneous coupled linear
equations for the dipole moments $\mathbf{p}_n$.

The resonant material response of the metal nanospheres is
captured in the frequency dependence of the polarizability
$\alpha(\omega)$. Within the Drude model, the polarizability of a
  nanosphere is   Lorentzian~\cite{pedrodevries}
\begin{equation}
\alpha_{\rm{Drude}}(\omega)=
4\pi\epsilon \frac{\omega_{SP}^2}{\omega_{SP}^2
-\omega^2-i\omega\gamma} a^3.
\end{equation}
Throughout this work, we use  values
$\gamma=8.25\cdot10^{14}$~s$^{-1}$, and $\omega_{SP}=4.64\cdot
10^{15}$~rad/s appropriate for silver particles in glass
($n=1.5$), as derived from tabulated optical constants in
Ref.~\cite{palik}. To maintain the energy balance between
extinction, scattering and absorption it is essential to include
   radiation damping   in the polarizability of each nanosphere  by setting the
polarizability to \cite{pedrodevries,weber,citrin,robicheaux}
$$
 {\alpha(\omega) }=\left[\frac{1}{\alpha_{\rm Drude}(\omega)}
-\frac{i}{6\pi\epsilon}\frac{\omega^3}{v^3}\right]^{-1}.
$$

\begin{figure}[t]
\centerline{\includegraphics[width=\figwidth]{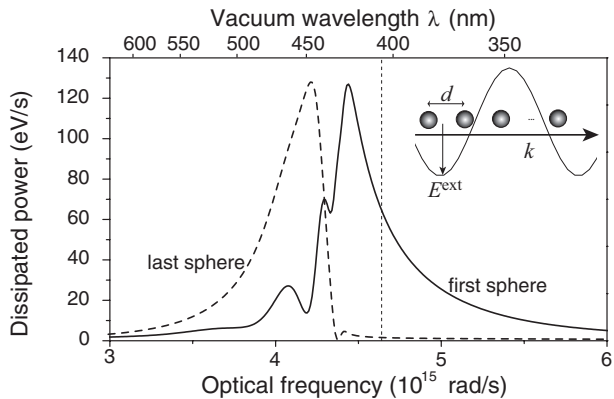}}
\caption{ Excitation by a plane wave   along the array causes a
localized   optical response in finite plasmon arrays (10 Ag
particles in glass, radius $30$~nm, $d=75$~nm). Depending on
driving frequency, the Ohmic dissipated power (at 1 V/m incident
field) is either concentrated on the first (solid curve), or last
sphere (dashed)   encountered by the incident beam. Vertical
dashes indicate the single particle resonance frequency.
 \label{fig1:asymmetricresponse}}
\end{figure}
The importance of retardation effects in the optical behavior of
metal nanoparticle chains is evident from
Figure~\ref{fig1:asymmetricresponse}(a), where we consider the
optical response of a chain of 10 nanoparticles of radius 30~nm
and spacing 75~nm 
illuminated by a plane wave incident along the array. We plot the
Ohmic dissipated power ($\propto |\mathbf{p}_n|^2$) for the first
and last sphere in the array as a function of the driving optical
frequency.  In agreement with results reported by Hernandez et
al.~\cite{robicheaux},  we find   a large asymmetry in the
response of the array, which can be tuned via the driving optical
frequency. For   frequencies below $4.3\cdot10^{15}$ rad/s the
backmost   nanosphere is preferentially excited (dashed line in
Fig.~\ref{fig1:asymmetricresponse}), i.e., the sphere that is
encountered last by the excitation beam. In contrast, this sphere
is hardly excited for higher frequencies, for which the frontmost
 sphere is strongly excited (solid curve).

\begin{figure}[tbh]
\includegraphics[width=\figwidth]{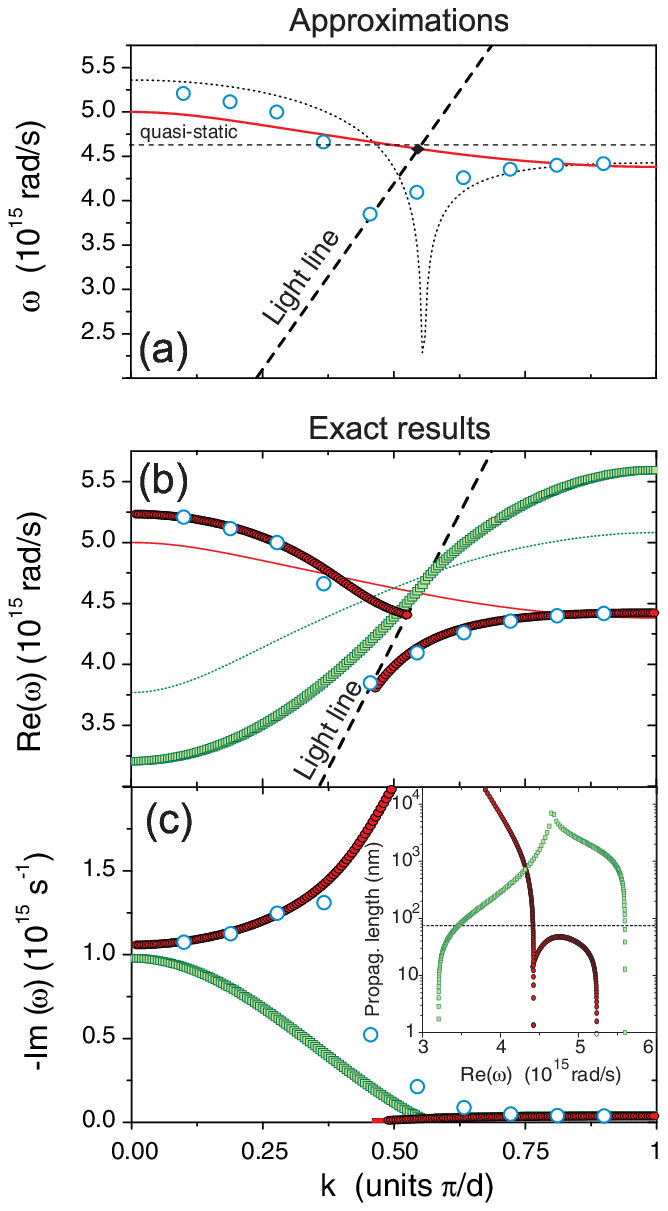}
\caption{(Color online) (A) The quasistatic dispersion (solid red
curve) for transverse modes intersects the light line (dashed)
slightly below the single particle resonance  (horizontal dashes).
Blue open circles show the   10 normal mode frequencies for an
array of 10 particles, according to Ref.~\cite{weber}, which agree
neither with the quasistatic dispersion, nor with the perturbative
result of Ref.~\cite{citrin} (dotted
curve).\label{fig:quasidisperse} (B) Red solid circles (green open
squares): real part of the frequency for roots of the exact
infinite chain dispersion  for transverse (longitudinal) modes.
For transverse modes two branches appear that anticross at the
light line (dashed). The result agrees well with the finite array
dispersion (blue open circles taken from (A)), but not with the
quasistatic approximation (thin curves, dotted for longitudinal
mode).  (C) Imaginary part of the frequency for the modes in (B).
The damping rate diverges near the light line for transverse
modes. Inset in (C): Propagation lengths (amplitude $1/e$ lengths)
versus frequency for transverse (red solid circles) and
longitudinal modes (green open squares). The horizontal dashed
line indicates the interparticle spacing
$d$.\label{fig:fullresult}}
\end{figure}

 Naively, one expects such strong coupling between the incident
wave and the plasmon chain to occur when the plasmon dispersion
relation intersects that of the embedding medium   ('light line'),
i.e., when the incident field is phase matched to the mode in the
array. For metal nanoparticle chains a quasistatic model is often
used~\cite{atwaterWGproposal,maierextinction,maierlossapl}, that
is found from Eq.~(\ref{eq:coupled}) for an infinite number of
dipoles, in the limit $\gamma=0, c\rightarrow \infty$ and in
absence of an external driving field. For the transverse modes
that the incident plane wave considered in
Figure~\ref{fig1:asymmetricresponse} can couple to
($\mathbf{p}_m\cdot \hat{\mathbf{r}}=0$), the quasistatic
dispersion relation reads
$$
\frac{\omega^2}{\omega_{SP}^2}=1+\frac{a^3}{d^3}\sum_{j=1}^\infty
\frac{2\cos j k d}{j^3},
$$
where $k$ is the wave vector. This dispersion relation is plotted
in Figure~\ref{fig:quasidisperse}(A) (red solid line).
 Evidently, the light line (dashed)
 intersects the quasistatic dispersion relation at a frequency of
 4.6$\cdot 10^{15}$ rad/s that does not correspond at all to the
 frequencies around 4.3$\cdot 10^{15}$~rad/s for which the
  complex response   in
Figure~\ref{fig1:asymmetricresponse} occurs.  Furthermore, our
exact calculations show that the asymmetric response  in
Fig.~\ref{fig1:asymmetricresponse} of arrays under plane wave
illumination shifts further to the red away from the plasmon
resonance as the spacing between particles is increased (data not
shown). This shift is in sharp contrast to the rapid $d^{-3}$
decrease of the bandwidth of the quasistatic dispersion relation
with increasing particle spacing, that is caused by  the strongly
reduced overlap of particles with the fields of their neighbors.
  We conclude that the
asymmetric response in Fig.~\ref{fig1:asymmetricresponse} points
at  the strong effect of retardation on the array dispersion.
Retardation affects the phases of individual nanospheres, and
thereby the constructive or destructive interference along the
chain.

Damping and radiation losses preclude the existence of a
dispersion relation which gives solutions of the form
$\mathbf{p}_m \propto e^{imkd}$  for real frequencies. However,
complex frequency solutions exist, corresponding to the dispersion
of damped modes. Recently, several authors have attempted to
determine the complex dispersion relation of plasmon chains. Weber
and Ford~\cite{weber} calculated the complex eigenfrequencies for
finite chains of $N$ dipoles, which are frequencies for which the
matrix coupling the dipoles in Eq.~(\ref{eq:coupled}) is singular.
A unique wave vector can be assigned to   these $N$ modes based on
their mode profile~\cite{weberemail}. The resulting set of
solutions is expected to be an accurate but discrete approximation
to the infinite chain dispersion relation, that includes effects
of retardation, and ohmic and radiation damping. Ref.~\cite{weber}
and open symbols in Fig.~\ref{fig:quasidisperse}(A) show that the
resulting dispersion, defined via the real part of the complex
frequencies, is dramatically different from the quasistatic
result, especially near the light line. Unfortunately, the
discrete sampling that is inherent in this approach  does not
allow to distinguish between the two distinct scenarios, in which
the disperion relation either has a polariton form with two
branches that anticross at the light line, or is a single
continuous dispersion relation.

To resolve this ambiguity, we calculate the infinite chain
dispersion by inserting $\mathbf{p}_m \propto e^{imkd}$ in
Eq.~(\ref{eq:coupled}), and setting the driving field to zero.
 Citrin~\cite{citrin} recently reported that
 the sums, which exhibit poor convergence for   frequencies
 with negative imaginary parts,  can be evaluated in terms of
polylogarithms~\cite{polylog} $\mathrm{Li}_n$, resulting in an
implicit dispersion relation for transverse modes of the form
\begin{equation}
0=1+\frac{\alpha(\omega) }{4\pi\epsilon d^3} \Sigma(\omega, k)
  \label{eq:selfenergy}
\end{equation}
with
\begin{eqnarray} \Sigma(\omega,k)&=& \left[\mathrm{Li}_3(e^{i
(\omega/v -k )d} ) +\mathrm{Li}_3(e^{i (\omega/v+k )d})  \right]\nonumber\\
&&- \frac{i\omega d }{v}\left[\mathrm{Li}_2(e^{i (\omega/v -k )d}
) +\mathrm{Li}_2(e^{i (\omega/v+k )d})\right] \nonumber \\ && +
\left(\frac{\omega d }{v}\right)^2\left[\mathrm{Li}_1(e^{i
(\omega/v -k )d} ) +\mathrm{Li}_1(e^{i (\omega/v+k )d})\right].
\label{eq:citrinsum}
\end{eqnarray}
In Ref.~\cite{citrin} this solution was   used   to approximate
the infinite chain dispersion   perturbatively by evaluating
$\Sigma(\omega,k)$ at the resonance frequency $\omega_{SP}$. The
result, plotted as the dotted line in
Fig.~\ref{fig:quasidisperse}(A), shows that this approximation
differs significantly from the quasistatic result and has a deep
minimum  where the light line crosses the quasistatic dispersion.
This minimum is due to the $1/r$ term in the dipole field, which
contributes the
  term
$ \left[\mathrm{Li}_1(e^{i (\omega/v -k )d} ) + \mathrm{Li}_1(e^{i
(\omega/v+k )d})\right] $ with   a logarithmic singularity at the
light line. As $\omega$ is far from $\omega_{SP}$ the validity of
this perturbative approach is limited.
Therefore we have numerically solved  the full dispersion relation
to find the complex  frequencies $\omega$ corresponding to real
wave vectors.

Figure~\ref{fig:fullresult}(B) (solid red dots) shows the
dispersion of the real part of the frequency for an infinite chain
of $a=30$ nm radius silver particles spaced by $d=75$~nm, embedded
in glass ($n=1.5$). First of all, our result confirms that the
deviations of the discrete finite chain  result in
Fig.~\ref{fig:quasidisperse}(A) from the quasistatic result for
infinite chains are not due to chain length effects, but to
retardation, radiation damping and ohmic damping. More
surprisingly, we find that the dispersion relation for transverse
excitations has two branches that anticross at the light line,
indicating polariton behavior of the coupled photon-plasmon chain
system. Regarding the upper dispersion branch, we note that this
branch does extend upwards along the light line, though our
numerical algorithm does not identify any clear roots as loss
rates become comparable to and exceed the optical frequency.
Indeed, the dispersion branch above the light line is accompanied
by huge damping, as gauged by the imaginary part of the frequency
plotted in Fig.~\ref{fig:fullresult}(C). Damping times for this
branch are shorter than about 1 optical cycle, and  radiative
losses diverge as the light line is approached.

 The existence of two dispersion
branches that display  an avoided crossing  is consistent with,
but could not be surmised from the finite chain result obtained by
Weber and Ford~\cite{weber}, and is not represented by the
perturbative treatment by Citrin~\cite{citrin} (dotted line in
Fig.~\ref{fig:quasidisperse}(A)). That two disconnected branches
appear may seem surprising in view of textbook treatments of,
e.g., the dispersion of surface plasmon polaritons. Damping causes
separate branches to merge when real frequencies but complex
k-vectors are considered~\cite{polaritonbook}. The dispersion of
complex frequencies for solutions with real wave vectors, however,
displays an   avoided crossing, also in the presence of damping. A
treatment in terms of complex frequencies, i.e. normal modes that
decay in time,  is the natural extension of the normal mode
analysis for finite arrays.

We now turn to the relevance of the exact dispersion relation
derived here for proposed applications of metal nanoparticle
chains as ultrasmall optical  waveguides.   Both the group
velocities   and the damping times in the present calculation are
profoundly different from those found in the quasistatic
approximation~\cite{atwaterWGproposal,maierlossapl,maierextinction,park}.
In Figure~\ref{fig:fullresult}(C, inset) we consider the
propagation length, defined as the product of the damping time in
Fig.~\ref{fig:fullresult}(C) and the group velocity that we derive
from Fig.~\ref{fig:fullresult}(B). Based on the quasistatic model,
the longest propagation lengths are expected near the center of
the Brillouin zone~\cite{maierlossapl,maierextinction}. Such
predictions as well as extrapolations from  ${k}=0$
measurements~\cite{maierextinction} need to be completely revised
due to the polariton splitting found here.  For transverse
excitations,   Fig.~\ref{fig:fullresult}(C) shows that modes above
the light line are strongly damped, with damping times comparable
to the optical period, and propagation lengths less than the
interparticle spacing (solid red dots). For modes just below the
light line ($k \gtrsim 0.3\pi/d$), however, no radiative losses
occur. Damping rates of around $10^{13}$~s$^{-1}$ and group
velocities around $0.3 c$ provide a frequency window for which
propagation lengths above $5~\mu$m are possible.

 Applications that require guiding  can benefit from
  longitudinal modes, for which the induced dipoles point along
the array. As was also found for finite arrays by Weber and Ford,
retardation effects cause the exact dispersion relation (open
squares in Fig.~\ref{fig:fullresult}(B))  to have a doubled
bandwidth compared to the quasistatic result (dotted curve). As
longitudinal modes do not couple to plane waves that propagate
along the array, the intersection with the light line is not
associated with diverging loss (Fig.~\ref{fig:fullresult}(C), open
squares) and no splitting of the dispersion relation occurs.  In
general, guiding benefits from the fact that the group velocity is
large over a much wider bandwidth. Below the light line a large
frequency window occurs for which the propagation length exceeds
1$~\mu$m.

In conclusion, we have presented the dispersion relation of
infinite metal nanoparticle chains, fully taking into account  the
effects of Ohmic damping, radiation damping and retardation. For
transverse modes strong coupling to plane waves propagating along
the array causes an unexpected splitting of the dispersion into
two branches, as common for polaritons. Propagation lengths just
above 1$~\mu$m for modes below the light line may allow plasmon
chains to be used as very small guiding structures for nanoscale
energy transport. With respect to the asymmetric optical response
in Figure~\ref{fig1:asymmetricresponse}, it appears that the local
excitation intensity set up by an incident plane wave, as gauged
by the Ohmic dissipation per sphere, can be used as a a sensitive
fingerprint of the   plasmon chain
 polariton anticrossing.
The sharply defined frequency at which the front-backward
asymmetry changes direction, and the peak absorption frequencies
for the front- and back sphere provide a measure for the location
and magnitude of the splitting in the dispersion relation. Such
steep frequency edges are sensitive to the refractive index of the
medium surrounding the array, and can be useful for, e.g.,
nanoscale sensing applications. Furthermore, the arrays allow to
create strongly localized energy distributions, that can be varied
by tuning the incident plane wave around the anticrossing in the
dispersion relation.  Such tunable localized energy distributions
are useful for, e.g., locally enhancing nonlinear interactions,
and nanoscale photolithography with visible wavelengths. Finally
we note that these phenomena should occur  for any chain of
coupled resonant dipoles, which also includes systems like quantum
dot arrays, or atomic systems~\cite{robicheaux2}.

We thank L.D. Noordam, W. H. Weber, G. W. Ford,  and L. Kuipers
for stimulating discussions. This work is part of the research
program of the ``Stichting voor Fundamenteel Onderzoek der Materie
(FOM),'' which is financially supported by the ``Nederlandse
Organisatie voor Wetenschappelijk Onderzoek (NWO).''


\begin{thebibliography}{10}

\bibitem{quinten}M. Quinten, A. Leitner, J. R. Krenn, and F. R. Aussenegg,
Opt. Lett. \textbf{23},  1331 (1998).




\bibitem{atwaterWGproposal} M. L. Brongersma, J. W. Hartman, and H. A.
Atwater, Phys. Rev. B \textbf{62}, R16356 (2000).

\bibitem{maierlossapl} S. A. Maier, P. G. Kik. and H. A. Atwater,
Appl. Phys. Lett \textbf{81}, 1714 (2002).




\bibitem{maierextinction}S. A. Maier, M. L. Brongersma, P. G.
Kik, and H. A. Atwater, Phys. Rev. B \textbf{65}, 193408 (2002).


\bibitem{atwaternatmat}S. A. Maier, P. G. Kik, H. A. Atwater, S.
Meltzer, E. Harel, B. E. Koel, and A. A. G. Requicha, Nature
Materials \textbf{2}, 229 (2003).


\bibitem{park} S. Y. Park and D. Stroud, Phys. Rev. B \textbf{69},
125418 (2004).


\bibitem{weber}W. H. Weber and G. W. Ford, Phys. Rev. B
\textbf{70}, 125429 (2004).

\bibitem{citrin} D. S. Citrin,  Opt Lett. in press 2005.


\bibitem{pedrodevries} P. de Vries, D. V. van Coevorden, and A. Lagendijk,
Rev. Mod. Phys. \textbf{70}, 447 (1998).
\bibitem{atomlattice}D.V. van Coevorden, R. Sprik, A. Tip, and A. Lagendijk,  Phys. Rev. Lett. \textbf{77},
2412  (1996).

\bibitem{robicheaux2}{F. Robicheaux, J. V. Hern\'andez, T.
Top\c{c}u, and L. D. Noordam, Phys. Rev. A \textbf{70}, 042703
(2004).}

\bibitem{robicheaux}J. V. Hern\'{a}ndez, L. D. Noordam, and F.
Robicheaux, J. Phys. Chem. B \textbf{109}, 15808 (2005).


\bibitem{palik} \textit{Handbook of Optical Constants of Solids}, edited
by E. D. Palik (Adacemic, Orlando, FL, 1985).




\bibitem{weberemail}W. H. Weber and G. W. Ford, personal communication.
 Contrary to the result quoted in Ref.~\cite{weber}, more than $N$ poles occur in the coupling
matrix for $N$ dipoles. Exactly $N$ poles can be traced to the $N$
modes in the quasistatic limit by their continuous dependence on
$c$ and $\gamma$.

\bibitem{polylog}{M. H. Lee, Phys. Rev. E \textbf{56}, 3909
(1997).}


\bibitem{polaritonbook}{H. Raether, \emph{Surface Plasmons},
(Springer, Berlin 1988).}


\end{thebibliography}
\end{document}